\documentclass[twocolumn,nofootinbib,showpacs,prd,superscriptaddress]{revtex4}

\usepackage{amsmath}
\usepackage{amsfonts}
\usepackage{amssymb}
\usepackage{graphicx}

\begin{document}

\title{Entanglement inside the cosmological apparent horizon}

\author{Salvatore Capozziello}
\affiliation{Dipartimento di Fisica, Universit\`a di Napoli ''Federico II'', Via Cinthia, Napoli, Italy.}
\affiliation{Istituto Nazionale di Fisica Nucleare (INFN), Sez. di Napoli, Via Cinthia, Napoli, Italy.}
\affiliation{Gran Sasso Science Institute (INFN), Via F. Crispi 7,  I-67100, L'Aquila, Italy.}

\author{Orlando Luongo}
\affiliation{Dipartimento di Fisica, Universit\`a di Napoli ''Federico II'', Via Cinthia, Napoli, Italy.}
\affiliation{Istituto Nazionale di Fisica Nucleare (INFN), Sez. di Napoli, Via Cinthia, Napoli, Italy.}
\affiliation{Instituto de Ciencias Nucleares, Universidad Nacional Autonoma de M\'exico (UNAM), Mexico.}

\begin{abstract}
Possible connections between quantum entanglement and cosmological eras are considered. In particular, assuming that two epochs are each other entangled, by measuring the entanglement degree, it is possible to recover  dynamical properties of the universe. In particular, the effects of dark energy could be due to the entanglement between states, since a negative pressure arises at late times. In this process, we choose as ruler to quantify the \emph{"entanglement weight"}, the so called \emph{negativity} of entanglement. It follows that a natural anti-gravitational effect occurs when the cosmological eras are  entangled. Thus, dark energy could be seen as a straightforward \emph{consequence} of entanglement. Specifically, our results  can be compared with observational data. In doing so, it is possible to show that a  pressureless term is recovered at a certain epoch dominating over dark energy and  ruling the structure formation. 
\end{abstract}


\date{\today}

\maketitle

\section{introduction}

Observational signatures of quantum gravity have been so far object of increasing interest \cite{kiefe}. Much efforts have been devoted both to investigate the quantum origin of structures \cite{kiefe2}, and to infer a possible quantum origin of the present universe dynamics \cite{tremendo}. In particular, Einstein's theory fails to be predictive at UV scales \cite{bingo,117www,Hartle:1986gn}, being overshadowed by  several shortcomings of the standard theory. Unfortunately, a self consistent description of quantum theory of gravity turns out to be so far unsuccessful. On the other hand,  approaches developing  \emph{quantum cosmology}  have been frequently investigated, in order to gain, at least, a quantum description of cosmological states \cite{kiefer3}. This point of view represent a {\it simplification} of the problem toward the full achievement of quantum gravity. In the Friedmann-Robertson-Walker (FRW) metric, the standard scheme of quantum cosmology leads to the introduction of the so-called {\it minisuperspace} (see \cite{odirev} for a recent review on the argument), where the wave function of the universe is obtained, reducing the degrees of freedom to a finite number. In other words, the main goal of quantum cosmology  is to  describe the observed universe starting from quantum theory that allows to set self-consistent initial conditions. So that, the observed cosmological parameters may be seen as the {\it average} result of a more general quantum descriptions. Significant arguments have been recently developed in favor of a quantum cosmology picture being independent of a full quantum gravity approach \cite{indep}. 

A  proposed point of view  relates quantum cosmology to quantum information theory \cite{Terno04}. In this picture, the notion of \emph{quantum entanglement} becomes a resource to describe the "\emph{interaction}" between cosmological quantum states. In this context, the guidelines rely on assuming quantum states for each cosmological epoch and  postulating the existence of entanglement between them. This \emph{entanglement ansatz} consists in choosing a multipartite quantum system where each part corresponds to a different era throughout the universe evolution \cite{tremendo,lettera}. The need of studying possible effects of informational theory comes from the idea of supposing that entanglement phenomena, between quantum states, could be responsible for the observed dynamics of the universe \cite{beh,sergei}. In addition, the role played by entanglement has been framed in robust theoretical and experimental tests \cite{tests}, assuming  entanglement as a true resource for theoretical physics, as well as energy, entropy, etc. \cite{goc}.

In this paper, we show that the  equation of state (EoS) of the universe can predict effects of dark energy and dark matter dominated era, respectively for redshift $z\ll1$ and $z\gg1$, according to the entanglement picture. Specifically, a detailed analysis of  cosmological entanglement between states can be achieved inside the {\it apparent horizon}, where  causality conditions are preserved \cite{go}. We propose a model where, starting from first principles, one can reconstruct the effects of  pressure $\mathcal P$. In addition, one can show that the pressure leads to anti-gravitational effects in  some regimes while, for $z\gg 1$, pressureless effects are achieved. In this approach,  dark energy is a dynamical consequence of the   entanglement between  cosmological states. 
In other words,  quantum terms are constrained by comparing them to the present cosmological observational bounds. A fairly good agreement is recovered, showing that the  observed  {\it coincidence problem } between dark matter and dark energy densities, in order of magnitude,  could be interpreted in view of  the  initial condition problem.

The paper is organized as follows. In Sec. \ref{sezdue}, we describe the  entanglement between cosmological states. We then highlight the basic demands of our approach, building up a self consistent framework for  quantum cosmological states. In Sec. \ref{seztre},  the role played by the so-called {\it negativity} of entanglement in cosmology is discussed. In Sec. \ref{sezquattro}, we describe how to relate entanglement  to negativity and develop the corresponding cosmological model, which derives from assuming the above hypotheses. In Sec. \ref{sezsei}, we describe the possible anti-gravitational effects due to entanglement and the fact that at early times, a pressureless term naturally arises.  Sec. \ref{sezsette} is devoted to conclusions and perspectives.


\section{Entanglement of cosmological states}\label{sezdue}

Let us  sketch the  hypothesis of assuming entanglement between cosmological states. In the  simplest picture, one can consider two cosmological epochs, characterized by some dynamical properties, such as inflation, reheating, recombination, and so forth. Afterwards, to each epoch, one associates a different Hilbert space. In a FRW universe, it naturally follows that fifferent epochs evolve changing their dynamical properties due to the existence of phase transitions \cite{coppa}. However, to guarantee that a perfect fluid exhibits a phase transition throughout the universe evolution, additional scalar fields or modifications of Einstein's gravity are requested \cite{campozzie}. Many approaches have been discussed in the literature, although none of them definitely clarify the nature of scalar fields or the origin of modified gravity. In other words, no final evidence for such phase transitions could be directly measured \cite{bull} also if, recently, indications emerged in this direction  \cite{bicep2}. Our aim here is to point out  a way to replace modifications of  cosmological standard model, by modeling  the  entanglement of cosmological states. In doing so, we assume that between two (or more) quantum cosmological states, the entanglement is responsible for the observed dynamical properties. This is a consequence of the fact that entanglement could enter the second law of thermodynamics \cite{mare}. A a possible approach to relate the entropy $\mathcal S$ to the mutual information, in the simplest case of isothermal processes, is to assume that
\begin{equation}\label{ciamocia}
\mathcal W \leq -\Delta \mathcal F+k_B T I_{\emph{ent}}\,,
\end{equation}
where $\mathcal W$ is the external net work on the system, $I_{\emph{ent}}$ the entanglement weight, here represented by the generalized mutual information content between the universe and the feedback controller. In the literature, such a mutual content is usually referred to as the QC-mutual information content. The acronym "QC" means that the universe cosmological states are quantum systems, while the observable are classical. In the picture of Eq. ($\ref{ciamocia}$), $T$ is the temperature and $\Delta \mathcal F$ is the Helmotz function, defined as
\begin{equation}\label{Helmotz}
\Delta \mathcal F=\rho \Delta\mathcal V-T\Delta\mathcal S\,,
\end{equation}
where the internal energy $U\equiv\rho\mathcal V$ and the total density $\rho$ have been considered according to \cite{mareale} and the net entropy reads ${\displaystyle \Delta \mathcal{ S} = \int \frac{d\Big[\Big(\mathcal P+\rho\Big)\mathcal V\Big]}{T}}$ (see \cite{iox}).
Thus, no  scalar field or  more complicated terms should  be accounted in the energy momentum tensor of general relativity according to this picture. To this end, the Hilbert space dimension is characterized by a  minimal set of observables, able to depict the universe dynamics at the time elapsing from the beginning to the end of phase transitions. Furthermore, to satisfy  the issue of determining a  minimal set of observables, we make use of the triangular cosmic relation. In fact, if no dissipative effects are taken into account, the fluid corresponding to the entanglement between states, could be considered as a perfect barotropic fluid \cite{fluid}. It follows that the EoS derived from entanglement is an additional counterpart to be included in the total energy-momentum  budget of the universe. In doing so, we consider two cosmological observables, the matter density $\Omega_{m}$ and the spatial curvature $\Omega_{k}$, associating to them a two-dimensional complex Hilbert space $\mathbb{C}^2$. The simplest (non-normalized) state takes the form
\begin{equation}
 |\phi\rangle\equiv \left(
\Omega_{m}+i\Omega_{k},\\
\Omega_{k}+i\Omega_{m}\right)^T\,,
\end{equation}
and the linear independent (non-normalized) vectors in $\mathbb{C}^2$ are
\begin{eqnarray}
|\tilde e_A\rangle&=&\left(
\Omega_{m}+i\Omega_{k},
i\Omega_{m}+\Omega_{k}\right)^T, \\
|\tilde e_B\rangle&=&\left(
\Omega_{m}-i\Omega_{k},
-i\Omega_{m}+\Omega_{k}\right)^T.
\end{eqnarray}
Such vectors represent the basis for building up the entire cosmological entangled state.

Following the Gram-Schmidt orthonormalization, we get
\begin{eqnarray}
|e_A\rangle&=&
N_A |\tilde e_A\rangle,\label{ortho1}\\
|e_B\rangle&=&
N_B (\Omega_m^2+\Omega_k^2) |\tilde e_B\rangle
+N_B 2i\Omega_m\Omega_k |\tilde e_A\rangle,
\label{ortho2}
\end{eqnarray}
where we defined the constants as $N_A=\frac{1}{\sqrt{2(\Omega_m^2+\Omega_k^2)}}$ and $N_B=\frac{1}{\sqrt{2(\Omega_m^2+\Omega_k^2)(\Omega_k^2-\Omega_m^2)^2}}$. Finally, the \emph{entangled states ansatz} requires that the universe evolves in an entangled state between the two epochs, i.e.
\begin{equation}
|\Psi\rangle = \alpha|e_A\rangle_1|e_B\rangle_2 + \beta|e_B\rangle_1|e_A\rangle_2\,,
\label{stato}
\end{equation}
where we made use of the simplest multipartite systems involving two epochs of the form
\begin{eqnarray}
 |e_A\rangle_{1} |e_A\rangle_{2},\;
  |e_A\rangle_{1} |e_B\rangle_{2},\;
   |e_B\rangle_{1} |e_A\rangle_{2},\;
    |e_B\rangle_{1} |e_B\rangle_{2},
 \label{basis}
\end{eqnarray}
which are associated to the total Hilbert space, i.e. $\mathbb{C}^2\otimes \mathbb{C}^2$. In our picture, $\alpha,\beta\in\mathbb{C}$ and satisfy the relation $|\alpha|^2+|\beta|^2=1$. For our purposes, without losing generality, we assume that $\alpha$ and $\beta$ are real and positive, so that $\beta=\sqrt{1-\alpha^2}$. This turns out to be adaptable to the cosmological context we are working in.

\section{Negativity as a measure of entanglement of cosmological states}\label{seztre}

We mentioned, in Sec. I, that the entanglement of quantum systems has been investigated as a resource for various typologies of information processes, giving insights in several fields \cite{hhh}. The development of experiments has deeply stimulated a great interest for the properties of  quantum systems both at microscopic and macroscopic levels. In particular, the problem of identification of possible quantum and classical correlations is of great interest \cite{gem}. Even though the relation between thermodynamics and entanglement has not been fully clarified so far, techniques for entanglement measurements are developing in order to quantify the quantum correlation of physical systems \cite{corre}. Examples are linear entropy, Von Neumann entropy, quantum discord, and so forth \cite{goc}. In this work, we mainly focus on the use of the so called \emph{negativity} of entanglement, in order to measure the possible effects on cosmological observables.

The negativity of entanglement is defined as follows \cite{negativity}
\begin{equation}\label{ns}
\mathcal{N}=2\sum_k \max(0,-\lambda_k)\,,
\end{equation}
where the sum is over the eigenvalues of the partially transposed density matrix. The first step is to evaluate the density matrix. It is easy to get
\begin{eqnarray}
\hat \rho = \left(
\begin{array}{cccc}
0 & 0 & 0 & 0\\
0 & \alpha^2 & \alpha\beta & 0\\
0 & \alpha\beta & \beta^2 & 0\\
0 & 0 & 0 & 0
\end{array}
\right).
\end{eqnarray}
Then, the partial transpose matrix with respect to the first system $R^{T_1}$, is
\begin{eqnarray}
\hat \rho^{T_1}=\left(
\begin{array}{cccc}
0 & 0 & 0 & \alpha\beta\\
0 & \alpha^2 & 0 & 0\\
0 & 0 & \beta^2 & 0\\
\alpha\beta & 0 & 0 & 0
\end{array}
\right).
\end{eqnarray}
The corresponding eigenvalues are written as $  \lambda_1 = \alpha^2,  \lambda_2 = \beta^2,  \lambda_3 = \alpha\beta$ and $  \lambda_4 = -\alpha\beta$.
It is easy to get the value of $\mathcal N$, as follows
\begin{equation}
\mathcal{N}(\hat\rho)=2\alpha\beta\,,
\label{negpsi}
\end{equation}
that is, in the case of cosmology,  constant throughout the universe evolution and depends on the two normalizing constants $\alpha$, $\beta$. The purpose of the next sections is to infer from Eq. ($\ref{negpsi}$) a general form of the cosmological model associated to a constant negativity, related to $\alpha$ and $\beta$. It is possible to  show that the classical observables of dark matter and dark energy can be tuned by considering the normalization quantum condition between $\alpha,\beta$, avoiding the coincidence problem \cite{coc}.

\section{Reconstruction of negativity through the cosmic horizon}\label{sezquattro}

In order to characterize $\mathcal N$, we need to define the cosmological volume where we consider our quantum states. In lieu of considering two separate spacetime regions, we could take into account the whole universe. Following the standard picture, one estimates   the entropy and temperature as functions of $\mathcal H$, i.e.
\begin{eqnarray}
  \mathcal S &\propto& \mathcal H^{-2}\,, \\
  \,\nonumber\\
  T &\propto& \mathcal H\,.
\end{eqnarray}
These relations give  match, in broad sense,   the thermodynamics coming from conformal field theory  to cosmology. This is analogous to the case of black hole thermodynamics, whose purview is to reproduce the first principle in terms of the horizon area. Recently, similar considerations have brought physicists to re-obtain the Friedmann equations in terms of the Cardy-Verlinde formula in the framework of conformal field theory. The philosophy is to recast the FRW metric in $n$ dimensions, as $ ds^2 = h_{ab}dx^adx^b +\tilde r^2 d\Omega_{n-1}^2$, with $\tilde r = a(t) r$ and $x^0=t$, $x^1=r$, and then to define the 2-dimensional metric $h_{ab} = {\rm diag} (-1, a^2/(1-\Omega_k r^2))$. Hence,  by employing $h^{ab} \partial_a \tilde r \partial _b \tilde r=0$,  it follows
\begin{equation}\label{cumba}
r_A = \frac{1}{\sqrt{\mathcal E^2 + \Omega_k (1+z)^2}}\,,
\end{equation}
where $\mathcal E\equiv\frac{\mathcal H}{\mathcal H_0}$. Note that Eq. ($\ref{cumba}$) represents the cosmological apparent horizon at all scales. We pick up Eq. ($\ref{cumba}$) as a general relation which exists for all redshift $z$. Other definitions, such as the photon causal distance $\tilde r$
\begin{equation}
\tilde r = \int_{t}^{t_0}{\frac{d\chi}{a(\chi)}}\,.
\end{equation}
 exists. It gives the distance that a photon travels from a light source at $r=r_0$ to our position at $r=0$.  Although such a definition preserves causality,  it works only for an accelerated and expanding universe. On contrary, the apparent horizon can be defined from $z=0$ to $z\gg1$. Recently, it has been proposed   that negativity can be expanded in terms of $a^{-2}$ at late times \cite{lettera}. At a first glance, this condition turns out to be predictive, since it fulfills the fact that $\mathcal N$ is a convex function for $z\ll1$. However, the corresponding cosmological model was not able to represent the universe dynamics in the range $z\gg1$. To this end, we are interested in extending the cosmological definition of $\mathcal N$, in order to achieve a more general relation between negativity and $r_A$. In doing so, let us notice that, in general, $\mathcal N=\mathcal F(r_A)$, where $\mathcal F(r_A)$ is a generic function of the apparent radius. Hence, we require that $\mathcal F$ has to satisfy the boundary conditions
\begin{description}\label{condizioni}
  \item[1] $\mathcal N(z=0)\sim0$\,,
  \item[2] $\mathcal N(z\rightarrow\infty)=1$\,,
  \item[3] $\frac{d\mathcal N}{dz}>0$\,,
\end{description}
which are conditions on $\mathcal N$, matching the cosmological and quantum definition of negativity. The first two conditions guarantee that, at early times, the entanglement is higher (condition 2), while at late times, the degree of entanglement is lower (condition 1). The meaning of condition 3 achieves the convexity of $\mathcal N$. Moreover, the Taylor expansion of $\mathcal N$, around the present epoch $(z\simeq 0)$, is immediately possible in order to relate such a quantity to cosmographic observables. To address the above conditions, we define
\begin{equation}\label{nra}
\mathcal N=1+\sigma\frac{dr_A}{dz}\,,
\end{equation}
that represents the simplest phenomenological parametrization of $\mathcal N$. It is easy to show that
\begin{equation}\label{kjsdh}
\frac{dr_A}{dz}=-\frac{1}{\mathcal H}\left(\frac{q+1}{1+z}\right)\,,
\end{equation}
where $q$ is the  deceleration parameter,
\begin{equation}\label{jd}
q\equiv-1+\frac{\dot{\mathcal H}}{\mathcal H^2}\,,
\end{equation}
where the dot represents the derivative with respect to the cosmic time. Since, at  $z\gg1$, the term $q\approx \frac{1}{2}$, and $\mathcal H\approx z^{3/2}$, we get
\begin{eqnarray}\label{kjhrd}
&&\lim_{z\rightarrow\infty}\frac{dr_A}{dz}=0\,,\\
\,\nonumber\\
&&\lim_{z\rightarrow 0}\frac{dr_A}{dz}\geq -1\,,
\end{eqnarray}
which satisfy the above conditions on $\mathcal N$. By employing Eq. ($\ref{negpsi}$), together with Eq. ($\ref{nra}$), we can recast $\Upsilon\equiv(2\alpha\beta-1)/\sigma$, and by using the Friedmann equations, i.e.
\begin{eqnarray}\label{ave2}
\mathcal H^2 &=& {8\pi G\over3}(\rho_m+\rho_{DE})\,,\\
\,\nonumber\\
\dot{\mathcal H} + \mathcal H^2&=&-{4\pi
G\over3}\left(3\mathcal P_{DE}+\rho_m+\rho_{DE}\right)\,,\label{ave22}
\end{eqnarray}
we are able to find out the corresponding dark energy density $\rho_{DE}$, emerged  from the entanglement between cosmological quantum states. It reads
\begin{equation}\label{hu}
\rho_{DE}=\rho_{cr} \frac{1-\Omega_k(1+z)^2\Upsilon^2\Big[z+\Upsilon^{-1}\left(1+\Omega_k-\Omega_m\right)^{-\frac{1}{2}}\Big]^2}{\Upsilon^2\Big[z+\Upsilon^{-1}\left(1+\Omega_k-\Omega_m\right)^{-\frac{1}{2}}\Big]^2}\,,
\end{equation}
where $\rho_{cr}=\frac{3H_{0}^{2}}{8\pi G}$ is the critical density. Note that Eq. ($\ref{hu}$) is  dependent on the scalar curvature $\Omega_k$. This property turns out to be negligible at recent times, while could be relevant during the transition between acceleration to deceleration regime.
In our picture, the cosmological constant may be interpreted as the lowest term of the dark energy density series, expanded around $z=0$. Moreover, the coincidence between $\Omega_m$ and $\Omega_\Lambda$ magnitudes could be re-interpreted as a problem due to the initial conditions of the evolving dark energy term. Framing it in other words, it turns out to be related to the dynamical initial conditions of the entanglement between quantum states.

\begin{table*}
\caption{{\footnotesize Values of the cosmographic set.}}
\begin{tabular}{c|c|c|c}
\hline\hline\hline
$\qquad${\small Range}$\qquad$   &  $\qquad$ $\mathcal N$ $\qquad$& $\qquad$ $\mathcal P$ $\qquad$ & $\qquad$ $|\Upsilon|$ $\qquad$ \\

\hline
$z\ll1$         & {$0.1\div 0.3$}    & {$<0$}    & {$0.45\div 1.2$}  \\[2.3ex]
\hline
$z\gg1$      & {$\sim 1$}      & {$\sim 0$}    & {$0.45\div 1.2$} \\[2.3ex]
\hline

\hline\hline\hline
\end{tabular}

{\footnotesize
Table of numerical results predicted by our model at small and high redshift regimes respectively. The range of mass is $\Omega_m=0.24\div0.30$. The parameter $\Upsilon$ remains unaltered throughout the universe evolution.}
\label{table:CS}
\end{table*}

\begin{figure}
\begin{center}\label{1dpdf}
\includegraphics[width=2.8in]{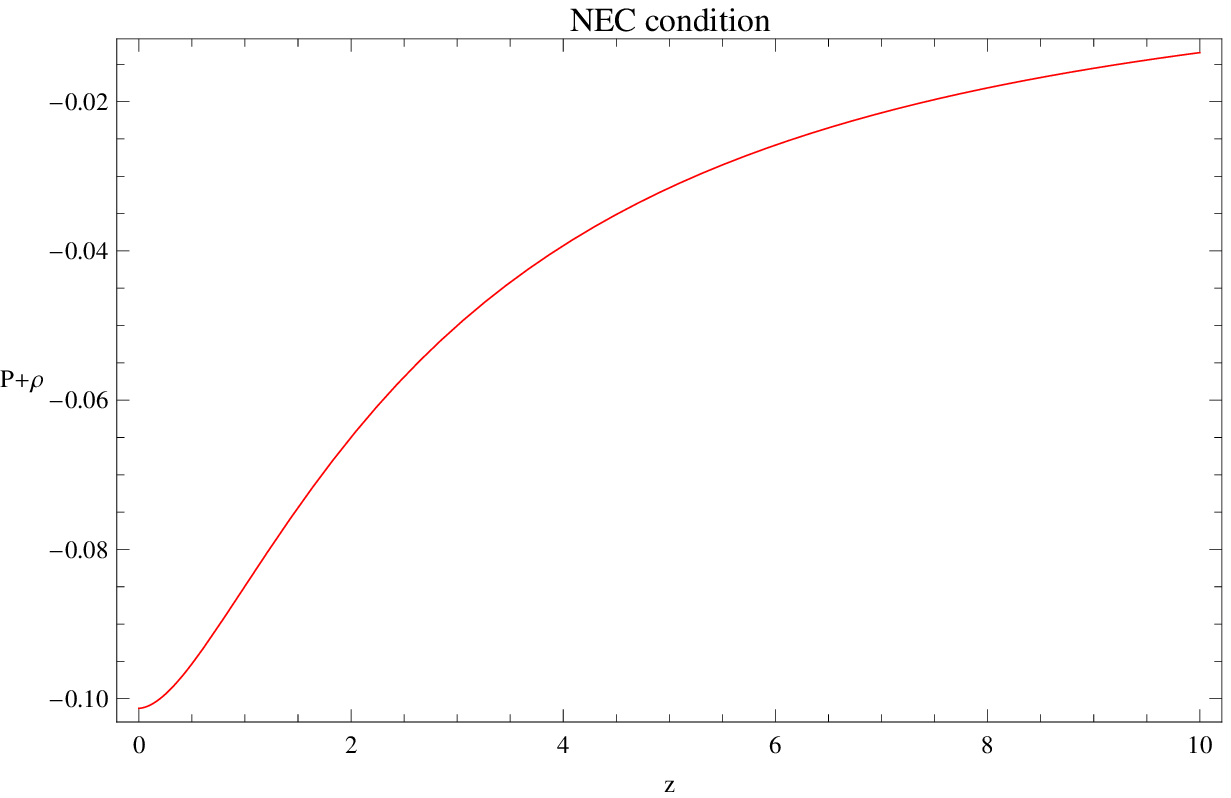}
\includegraphics[width=2.8in]{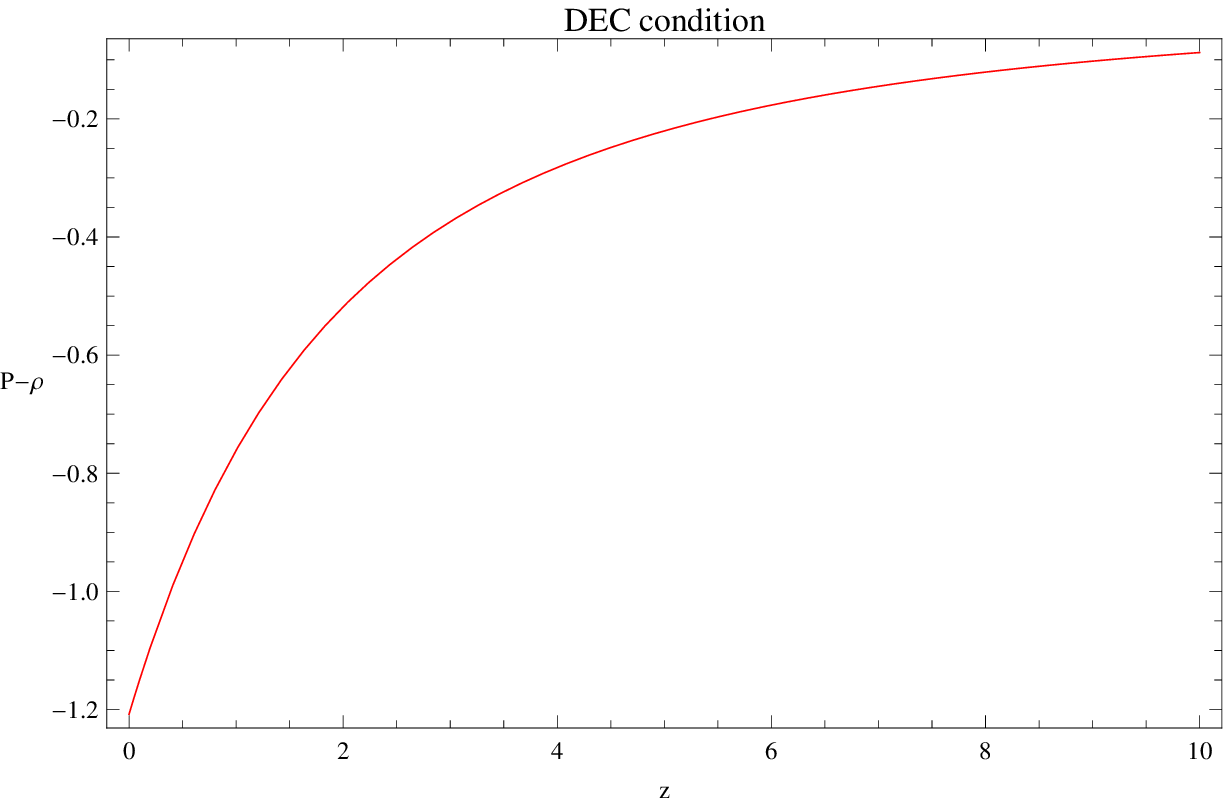}
\includegraphics[width=2.8in]{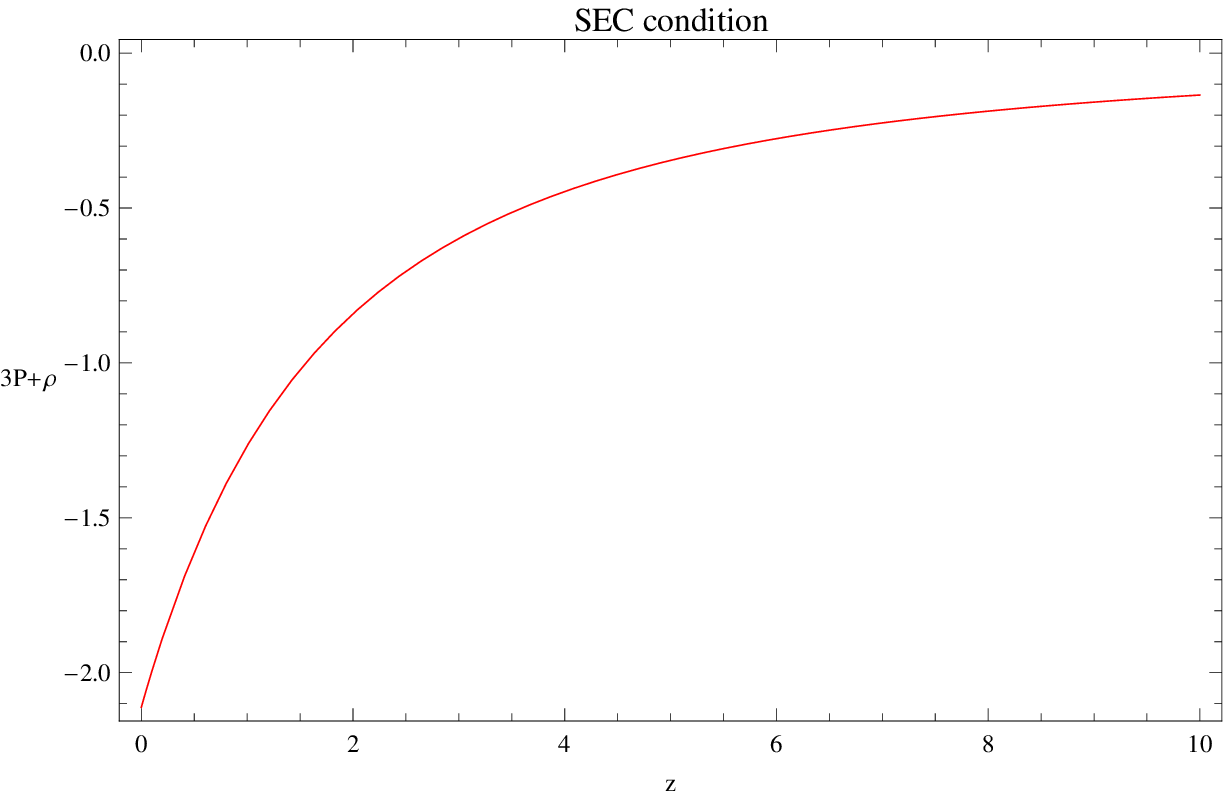}
\caption{Representation of the Energy Conditions for the particular particular value $\Upsilon=1.37$, with the WMAP 7-year value $\Omega_m=0.274$ \cite{wmap}.}
\end{center}
\end{figure}

\section{Repulsive effects and negative pressure}\label{sezsei}

Let us consider now Eqs. (\ref{ave2}) and (\ref{ave22}), and  write the pressure and density in the following form
\begin{equation}
\frac{\mathcal P}{\rho} = \frac{1}{3} (2 q -1)\,.
\end{equation}
Thus, by defining the dark energy EoS as $\omega\equiv\frac{\mathcal P}{\rho_{DE}}$, we infer the continuity equation of $\omega$ in terms of $\mathcal E\equiv\frac{\mathcal H}{\mathcal{ H}_0}$, being 
\begin{equation}\label{dkjb}
\omega=-1-\frac{2}{3}\left(\frac{d\ln \mathcal E}{d\ln a}\right)\,,
\end{equation}
where we used the definition  $a\equiv\frac{1}{1+z}$. It is easy to get the form of the pressure
\begin{equation}\label{pressure}
\mathcal P =-\frac{3\left(1-2\Omega_m\right)+2\Upsilon\left(1-2\Omega_m\right)^{\frac{3}{2}}+5\Upsilon z\left(1-2\Omega_m\right)^{\frac{3}{2}}}{3\Big[1+\Upsilon z\left(1-2\Omega_m\right)^\frac{1}{2}\Big]^3}\,.
\end{equation}
The dark energy effect is  realized for  $\mathcal P<0$. On the other hand, a standard barotropic fluid needs that $\mathcal P>0$; however, by considering  the entanglement negativity   that gives  $\alpha<1$ and $\beta<1$, it is easy to provide ranges where the pressure exhibits negative behaviors. To clarify this statement, let us focus on the energy conditions (ECs) for a  barotropic fluids \cite{ellis}, i.e.
\begin{eqnarray}\label{ciao}
  T_{\mu\nu}n^\mu n^\nu&\geq&0\,, \\
      T_{\mu\nu}n^\mu n^\nu&\geq&0\quad {\rm and} \quad T_{\mu\nu}T^{\nu}_{\delta}t^\mu t^\delta\leq0\,, \\
    T_{\mu\nu}n^\mu n^\nu&\geq&\frac{1}{2}T_{\gamma}^{\gamma}t^{\delta}t_{\delta}\,,
\end{eqnarray}
which have the advantage to be coordinate-invariant constraints on the energy-momentum tensor. These equations represent the most relevant assumptions for describing the regions of repulsive effects (see also \cite{lobo} for a discussion on ECs in modified theories of gravity). They are respectively the null (NEC), the strong (SEC), and the dominant energy condition (DEC). These relations are valid for  timelike vectors $t^\mu$. Physically, it follows that the first of these constraints, the NEC, is related to the stability of fluids, \cite{ene2}. The second, the DEC, determines a lower bound on EoS of dark energy \cite{ene3}, while the third can be  violated, in order to accelerate the universe. Violations of  SEC are common for a wide variety of dark energy models \cite{ene6}.

For a standard perfect fluid regime, we get
\begin{eqnarray}
  P+\rho \geq 0\,, \label{aaaa1}\\
  \rho>0\,,\quad -\rho\leq P\leq \rho\,,\label{aaaa2}\\
  \rho+3P\geq 0\,.\label{aaaa3}
\end{eqnarray}
On the other hand, the condition $\rho+3P\leq 0$ gives rise to the accelerated behavior of the cosmic fluid. 

In Figs.1, the ECs are represented  as functions of the redshift $z$ for a given value of $\Upsilon$ . We found the regions of repulsive effects,  coinciding  with the accelerating present epoch,  as confirmed by observations. In addition, for $z\gg1$, a pressureless behavior is recovered, that is the matter dominated epoch is achieved  assuming the  entanglement between cosmological quantum states. In particular, for a perfect fluid, the corresponding perturbation relation is
\begin{equation}\label{er}
\delta \mathcal{P}=c_{a}^{2}\delta\rho+\tilde{\sigma}\delta \mathcal S\,.
\end{equation}
In the case of adiabatic perturbations, the entropy perturbations $\tilde{\sigma}$can be neglected, and the sound speed definition depends only on the adiabatic perturbations of pressure. Thus, defining $c_s$ as the general sound speed, we have that $c_s=c_a$ only in the region where the DEC condition is preserved. In such a case, the $c_a$ definition reads
\begin{equation}\label{ass}
c_a^2 \equiv \left(\frac{\partial \mathcal P}{\partial \rho}\right)_S=\frac{\dot{\mathcal P}}{\dot \rho}\,,
\end{equation}
that is related to the structure formation. In fact, it is well known that structures can form at all scales, if the Jeans length, i.e. the sound speed, is negligibly small. For our model, the general expression for the sound speed reads
\begin{equation}\label{sound}
c_s^2=-\frac{2+3\Upsilon\left(1-2\Omega_m\right)^{\frac{1}{2}}+5\Upsilon z\left(1-2\Omega_m\right)^{\frac{1}{2}}}{3\Big[1+\Upsilon z\left(1-2\Omega_m\right)^\frac{1}{2}\Big]}\,.
\end{equation}
This expression can  be negative, because not related to the adiabatic sound speed for all $z$. Results coming from the values of $c_s$, $\mathcal P$ and $\rho_{DE}$ are reported in Tab. I, where we show a fairly good agreement with current cosmological data. In particular, at $z\gg1$ $c_s\rightarrow 0$ and then the matter dominated  behavior is recovered.

\section{Discussion and Conclusions}\label{sezsette}

In this paper, we showed that a possible correspondence between the apparent horizon and the so called negativity of entanglement may occur, when two or more than two quantum cosmological states are entangled to each other. Hence,  entanglement could represent the \emph{source of missing energy budget} associated to dark energy. In other words, the fluid which drives the universe dynamics today may be composed by the standard  pressureless matter term, the scalar curvature density, the radiation term and a fluid whose origin  is related to the entanglement mechanism. We showed that the dark energy effects are well mimicked by such an \emph{entanglement fluid}, which reproduces both present (accelerating) and early (matter dominated) eras. In particular, we found that entanglement is able to start to accelerate the universe in the regime $z\ll1$, fixing as initial conditions, the value of the cosmological constant today. This alleviates the so called coincidence problem. Moreover, the fluid derived from entanglement, is even able to represent a fairly good description of the universe dynamics at early times. To prove these facts, we inferred the ranges where the pressure of the \emph{entanglement fluid} becomes negative and the ranges where it changes sign, showing a phase transition. To achieve this result, we investigated the  energy conditions of the energy momentum tensor, adapted for a perfect fluid. In addition, the standard picture of structure formations is recovered, since no significative departures from the speed of sound and Jeans length occur. 

 An important point has to be clarified according  to the possible  anti-gravitational 
effects  taken into account in Sec.V.  As discussed in details in \cite{antigravity},  antigravity does not appear  in the early 
universe due to conflict with dynamical equations. However,  the further degrees of freedom related to $f(R)$ gravity can effectively act as scalar fields capable of leading accelerated behaviors both in inflation and in dark energy eras. Here, the situation is very similar since entanglement gives rise to effective negative pressure whose net effect is an apparent antigravity. The  result is  the violation of the energy conditions. 
Future efforts will be devoted to investigate the consequences of more complex multipartite quantum states, in the context of observable cosmology.

\acknowledgements
SC is  supported by INFN (iniziative specifiche QGSKY, TEONGRAV). OL is supported by the European PONa3 00038F1 KM3NET (INFN) Project.


\begin{thebibliography}{99}

\bibitem{kiefe}
C. Kiefer, M. Kraemer, Phys. Rev. Lett. 108, 021301, (2012); Int. J. Mod. Phys. D, 21, 1241001, (2012).

\bibitem{kiefe2}
C. Kiefer, Class. Quant. Grav., 4, 1369, (1987); C. Lammerzahl, Phys. Lett. A, 203, 12, (1995).

\bibitem{tremendo}
S. Capozziello, O. Luongo, Entropy, {\bf 13}, 2, 528, (2011).

\bibitem{bingo}%
J. B. Hartle, in \emph{The Quantum Structure of Space and Time}, Proceedings of the $23^{rd}$ Solvay Conference, ArXiv[gr-qc]:0602013.

\bibitem{117www}%
J. J. Halliwell, J. B. Hartle, Phys. Rev. D, 41, 1815, (1990).

\bibitem{Hartle:1986gn}%
J. B. Hartle, {\emph Gravitation in Astrophysics},  Proceedings of Cargese Conference, (1986).

\bibitem{kiefer3}
C. Kiefer, B. Sandhoefer, \emph{Quantum Cosmology}, arXiv:0804.0672, (2008).

\bibitem{odirev}
S. Capozziello, M. De Laurentis, S. D. Odintsov, Eur. Phys. J. C, 72, 2068, (2012).

\bibitem{indep}
C. Kiefer, {\emph Quantum Gravity}, Oxford Univ. Press, Oxford, (2007).

\bibitem{Terno04}
A. Peres, D. R. Terno, Rev. Mod. Phys., 76, 93, (2004).

\bibitem{lettera}
S. Capozziello, O. Luongo, S. Mancini,  Phys. Lett. A 377, 1061, (2013).

\bibitem{beh}
M. Aspachs, G. Adesso, I. Fuentes, Phys. Rev. Lett., 105, 151301, (2010); P. M. Alsing, I. Fuentes, Class. Quant. Grav., 29, (2012); N. Friis, D. E. Bruschi, J. Louko, I. Fuentes, Phys. Rev. D, 85, 081701, (2012).

\bibitem{sergei}
S. Nojiri and S.D. Odintsov, Phys. Rept. 505, 59  (2011).


\bibitem{tests}
S. P. Walborn et al., Nature, 440, 1022, (2006); C. Schmid et al., Phys. Rev. lett., 101, 260505, (2008); G. Vidal, J. Mod. Opt., 47, 355, (2000).

\bibitem{goc}
M. B. Plenio, S. Virmani, Quant. Inf. Comput., 7, 1-51, (2007).

\bibitem{go}
R. G. Cai, S. P. Kim, JHEP., 0502, 050, (2005); R. G. Cai, L. M. Cao, Y. P. Hu, Class. Quant. Grav., 26, 155018, (2009).

\bibitem{coppa}
J. E. Copeland, M. Sami, S. Tsujikawa, Int. J. Mod. Phys. D, 15, 1753-1936, (2006).

\bibitem{campozzie}
K. Bamba, S. Capozziello, S. Nojiri, S. D. Odintsov, Astroph. and Sp. Sci., 342, 155-228,  (2012); S. Capozziello and M. De Laurentis, Phys. Rept. 509, 167, (2011);
S. Capozziello, M. De Laurentis, O. Luongo, A.C. Ruggeri, Galaxies, 1, 216 (2013).

\bibitem{bull}
M. Li, X.D. Li, S. Wang, Y. Wang, M. Li, X.-D. Li, Shuang Wang, Yi Wang, Commun. Theor. Phys. 56, 525-604, (2011).

\bibitem{bicep2}
P. A. R. Ade et al. [BICEP2 Collaboration], arXiv:1403.3985 [astro-ph.CO], (2014).

\bibitem{mare}
T. D. Kieu, Phys. Rev. Lett. 93, 140403, (2004); K. Maruyama, F. Morikoshi, V. Vedral, Phys. Rev. A, 71, 012108, (2005); J. Oppenheim, M. Horodecki, P. Horodecki, R. Horodecki, Phys. Rev. Lett., 89, 180402, (2002); B. Piechocinska, Phys. Rev. A, 61, 062314, (2000); M. O. Scully, Phys. Rev. Lett. 87, 220601,  (2001).

\bibitem{mareale}
T. Sagawa, M. Ueda, Phys. Rev. Lett., 100, 080403, (2008).

\bibitem{iox}
O. Luongo, H. Quevedo, Gen. Rel. Grav. 46, 1649,  (2014).

\bibitem{fluid}
E. V. Linder, R. J. Scherrer, Phys. Rev. D, 80, 023008, (2009).

\bibitem{hhh}
R. Horodecki, P. Horodecki, M. Horodecki, K. Horodecki, Rev. Mod. Phys., 81, 865-942, (2009).

\bibitem{gem}
A. Matzkin, AIP Conf. Proc., 1384, 27, (2011).

\bibitem{corre}
J. Sirker, J. Stat. Mech. P12012, (2012); V. Vedral, M. B. Plenio, M. A. Rippin, P. L. Knight,
Phys. Rev. Lett., 78, 2275, (1997); M. B. Plenio, S. Virmani, Quant. Inf.  Comp., 7, 1, (2007).

\bibitem{negativity}
L. Amico, R. Fazio, A. Osterloh, V. Vedral, Rev. Mod. Phys., 80, 517, (2008); J. Eisert, M. Cramer, M. B. Plenio, Rev. Mod. Phys., 82, 277, (2010); P. Calabrese, J. Cardy, B. Doyon, J. Phys. A, 42, 500301, (2009).

\bibitem{coc}
N. Arkani-Hamed, L. J. Hall, C. Kolda, H. Murayama, Phys. Rev. Lett., 85, 4434, (2000).

\bibitem{wmap}
E. Komatsu {\it et.al.}, Astrophys. J. Supp. {\bf 192}, 18, (2011).

\bibitem{ellis}
S.W. Hawking and G.F.R. Ellis, {\it The Large scale structure of space-time}, Cambridge Univ. Press, Cambridge (1973).

\bibitem{lobo}
S. Capozziello, F.S.N. Lobo, J.P. Mimoso, Phys. Lett. B 730, 280 (2014).

\bibitem{ene2}
S. Dubovsky, T. Gregoire, A. Nicolis, R. Rattazzi, J. High Energy Phys., 3, 25, (2006); M. P. Lima, S. Vitenti, M. J. Reboucas, Phys. Rev. D, 77, 083518, (2008).

\bibitem{ene3}
R. V. Buniy, S. D. H. Hsu, B. M. Murray, Phys. Rev. D, 74, 063518, (2006).

\bibitem{ene6}
S. M. Carroll, M. Hoffman, M. Trodden, Phys. Rev. D, 68, 023509, (2003); C.-J. Wu, C. Ma, T.-J. Zhang, ApJ, 753, 97, (2012); P. Schuecker, R. R. Caldwell, H. Bohringer, C. A. Collins, L. Guzzo, N.  Weinberg, AA, 402, 53, (2003); K. Lake, Class. Quant. Grav., 21, L129, (2004).

\bibitem{antigravity}
K.  Bamba, S. Nojiri, S. D. Odintsov, D.  Saez-Gomez, Phys. Lett. B 730, 136  (2014).

\end{thebibliography}
\end{document}